%% file: main.tex
\documentclass[twoside]{article}

\usepackage{lipsum} 

\usepackage[sc]{mathpazo} 
\usepackage[T1]{fontenc} 
\usepackage{microtype} 

\usepackage[hmarginratio=1:1,top=32mm,columnsep=20pt]{geometry} 
\usepackage{multicol} 
\usepackage[hang, small,labelfont=bf,up,textfont=it,up]{caption} 
\usepackage{booktabs} 
\usepackage{float} 
\usepackage{hyperref} 

\usepackage{lettrine} 
\usepackage{paralist} 

\usepackage{graphicx} 
\usepackage{subcaption}
\usepackage{booktabs}

\usepackage{abstract} 

\usepackage{titlesec} 
\renewcommand\thesection{\arabic{section}} 
\renewcommand\thesubsection{\thesection.\arabic{subsection}} 
\titleformat{\section}[block]{\large\scshape\centering}{\thesection.}{1em}{} 
\titleformat{\subsection}[block]{\large}{\thesubsection.}{1em}{} 

\usepackage{fancyhdr} 
\title{\vspace{-15mm}\fontsize{24pt}{10pt}\selectfont\textbf{Using RDMA for Lock Management}} 

\author{
\normalsize{} 
\large
\textsc{Yeounoh Chung} \\
\textsc{Erfan Zamanian} \\[2mm]
\normalsize{\{yeounoh, erfanz\}@cs.brown.edu} 
\and
\normalsize{Supervised by:} \\[1mm]
\textsc{John Meehan} \\
\textsc{Stan Zdonik} \\[2mm]
\normalsize{\{john, sbz\}@cs.brown.edu}
\vspace{5mm}
}

\date{}

\begin{document}

\maketitle 

\thispagestyle{fancy} 



\begin{abstract}
In this work, we aim to evaluate different Distributed Lock Management service designs with Remote Direct Memory Access (RDMA). In specific, we implement and evaluate the centralized and the RDMA-enabled lock manager designs for fast network settings. Experimental results confirms a couple of hypotheses. First, in the traditional centralized lock manager design, CPU is the bottleneck and bypassing CPU on client-to-server communication using RDMA results in better lock service perofrmance. Second, different lock manager designs with RDMA in consideration result in even better performance; we need to re-design lock management system for RDMA and fast networks.
\end{abstract}


\begin{multicols}{2} 

\section{Introduction}
\label{Introduction}
Lock management is a critical component of many distributed systems, such as databases and file systems, in which shared resources are accessed by many applications across the network. A lock manager provides advisory locking services to these higher level applications and therefore is used to ensure serialization of access to shared resources (for example databases need this property to ensure the isolation between concurrent transactions).
In a single-sited lock managers, clients have to contact the centralized lock service before accessing or moving any data item. In such systems, reasoning about starvation and deadlocks is relatively straightforward, and is handled by the central lock server. They are rarely used in distributed systems, however, due to two main reasons. First, as all requests have to be handled by a single machine, the central lock manager becomes the bottleneck and therefore such systems do not scale with the number of clients. Second, loosely-coupled distributed systems are usually deployed on commodity clusters, where failures are quite common. A failure or delay in the central lock manager could bring the entire system down.
In this work, we aim to evaluate different Distributed Lock Management service designs with Remote Direct Memory Access (RDMA). In specific, we implement and evaluate the centralized and the RDMA-enabled lock manager designs for fast network settings. Our hypothesis is that the centralized lock manager is the bottleneck and bypassing CPU on client-to-server communication using RDMA results in better lock service performance; furthermore, different lock manager designs with RDMA in consideration result in even better performance.
The key idea is to have clients take part in lock management via RDMA, and our experimental results show that the new lock manager designs along this line can outperform the traditional centralized lock manager by an order of magnitude.
In the following sections, we first describe RDMA features and basics, describe different lock manager designs and their performances over a fast network. Finally, we highlight the comparison results in section 5. And we discuss any related work to this project in section 6 and conclude in section 7.


\section{Remote Direct Memory Access}
\label{background:rdma}
Remote Direct Memory Access (RDMA) is an emerging technology in computer networks which allows machines send or receive messages directly to each other's memory.
Invoking this process does not involve any system call and context switch from the user space to the kernel space.
Besides, throughout the process, kernel of either side is not involved.
As most operating systems impose high overheads for interrupt processing, network protocol stacks, and context switching, avoiding them significantly helps decrease the latency.

RDMA programming model is based on the concept of \emph{verbs}, which allows for two communication models: one-sided communication and two-sided communication.

\textbf{One-sided}, or \textbf{memory semantic} verbs, are those verbs which are performed without any knowledge of the remote side.
RDMA READ, WRITE, and atomic operations, such as Compare and Swap (CS), and Fetch and Add (FA) are one-sided operations.
The active side submits the verb, while the passive side is completely unaware of the process.
Both active and passive sides must register the memory region to be able to access it via RDMA.
The passive side's RNIC directly writes/fetches the desired data using an DMA operation from local memory.

\textbf{Two-sided}, or \textbf{channel semantic} verbs, such as SEND and RECEIVE, require both sites to involve in the communication process.
The payload of the SEND is written to the memory region specified by a corresponding RECEIVE which must be posted by the receiver before the sender actually sends its request.
Therefore, the sender's RNIC does not have to register the remote memory region before performing the operation.

While two-sided verbs are workflow-wise similar to socket programming semantics, namely read() and write(), leveraging the one-sided verbs, such as READ and WRITE, requires a dramatic change in the programming model. 

RDMA uses the concept of queue pairs for connections.
Application posts verbs to the send queue, which has a corresponding receive queue on the receive side, so the name queue pair.
Once RNIC performs the requested memory access at the remote side, it pushes a completion event on a corresponding completion queue, which can notify the sender about the completion of the task.
All queues are maintained inside RNIC.


\begin{figure*}[t!]
\begin{subfigure}{.5\textwidth}
  \centering
  \includegraphics[width=.8\linewidth]{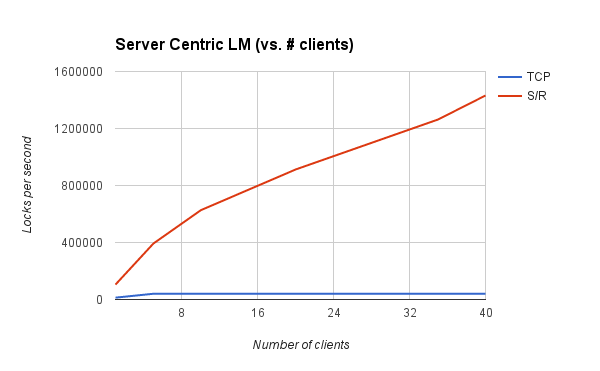}
  \caption{vs. number of client}
\end{subfigure}%
\begin{subfigure}{.5\textwidth}
  \centering
  \includegraphics[width=.8\linewidth]{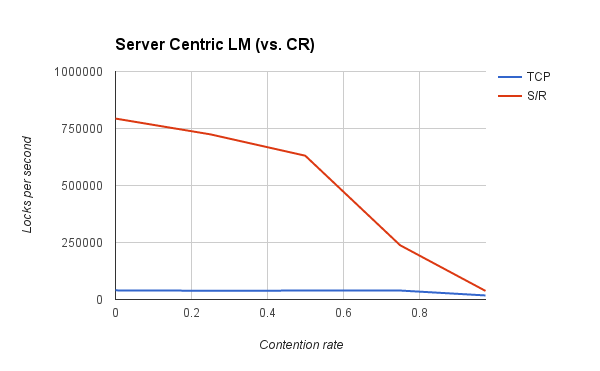}
  \caption{vs. contention rate}
\end{subfigure}
\caption{Throughput of centralized lock manager with RDMA send/receive verbs}
\label{fig:server_centric_result}
\end{figure*}
 
\input{server_centric.tex}


\section{Pure Client Centric Design}
\label{client_DLM}

\begin{figure*}
    \centering
    \includegraphics[scale=0.4]{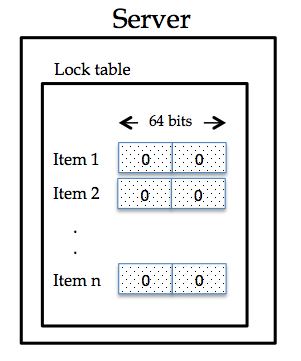}
    \caption{Server's memory layout for client centric design}
    \label{fig:pure-design}
\end{figure*}
One problem with server centric designs is that server could become the bottleneck as the number of clients grows.
In order to support concurrent requests, servers are usually implemented as a multi-threaded program.
In such cases, accessing or modifying the state of locks requires using mutexes, semaphores or other similar concurrency control mechanisms, which is expensive.
Given that the logic of a lock manager is relatively simple (lock and unlock an item), RDMA could be leveraged to allow the clients directly participate in lock management.

We started out with an extreme design, where RDMA is leveraged as much as possible.
The basic idea here is that server is completely oblivious of lock requests, and serves merely as the lock holder, while clients directly compete for lock acquisition.
As such, clients try to ``register'' themselves as the shared or exclusive owner of the lock by issuing RDMA atomic operations.
In the following, we will explain the details of our algorithm.
Lock states are kept inside the \emph{lock table}, which resides in the main memory, whose structure is shown in Figure \ref{fig:pure-design}.
Entry \textsc{i} of this table represents the status of lock \textsc{i} as a 64-bit memory region, and is divided to two 32-bit parts.
the most significant 32 bits represent the client ID which is currently owning the \emph{exclusive} lock. 
The least significant 32 bits represent the number of requests for the shared lock.

In the following subsections, we outline the procedure for acquiring and releasing locks.

\subsection{Exclusive Lock Acquisition}
As a client can only acquire exclusive lock when nobody else is holding the lock (neither in shared mode nor exclusive), the procedure must be only successful if these requirements are fulfilled.
Therefore, the client submits an atomic \emph{Compare-and-Swap} to the server with the following parameters: (0|0) as the expected value, and (client\_id|0) as the swap value.
The first number of each tuple is the most significant 32 bits, and the second one is the least significant one.
Note that atomic operations always return the old value of the asked memory regions, regardless of whether or not they succeeded.
Therefore, the client can know whether its CAS request was successful or not, by comparing the return value to (0|0).
Any value other than this means that the lock is not acquired.

If it was successful, the client does not need to do anything else.
If not, then it backs off for a pre-determined time, and try again with the same parameters.
We found that the back-off should of 0 seconds gives the best result.

\subsection{Shared Lock Acquisition}
\label{pure:shared_acquisition}
Shared lock can be acquired when nobody is holding the lock, or if all the owners are also in shared mode.
Therefore, the algorithm must make sure that the shared lock is not granted only when there is an exclusive owner.
Therefore, the client sens an atomic \emph{Fetch-and-Add} with increment value equals one to the server.
Note that atomic FA is always successful and adds the specified value to the memory region.

Again, the return value of this operation determines whether the lock is granted or not.
If the exclusive part of returned value (the most significant 32 bits) is zero, then the client can go ahead with the shared lock.
If not, then the lock must be exclusively owned by another client.
In this case, similar to exclusive lock, the client backs off for a specific amount of time, and then retry.
However, retrying is not be done by submitting another FA operation, since calling FA again will cause the lock status to get incremented once more.
Therefore, the consequent retries will be done via RDMA READ of the exclusive part (only 32 bits).
If the client finds that the lock is not in exclusive mode, it can proceed with the shared lock.

\subsection{Exclusive Lock Release}
In order to release an exclusive lock, the client only needs to clear the exclusive part of the lock status.
This can be easily done by RDMA WRITE.
More specifically, the client WRITEs zero to the exclusive part of the lock.

\subsection{Shared Lock Release}
As we have seen in Section \ref{pure:shared_acquisition}, acquiring shared lock involves atomically incrementing the shared part of the lock status.
Consequently, releasing the shared lock involves decrementing the same region by one.


\begin{figure*}[t!]
\begin{subfigure}{.5\textwidth}
  \centering
  \includegraphics[width=.8\linewidth]{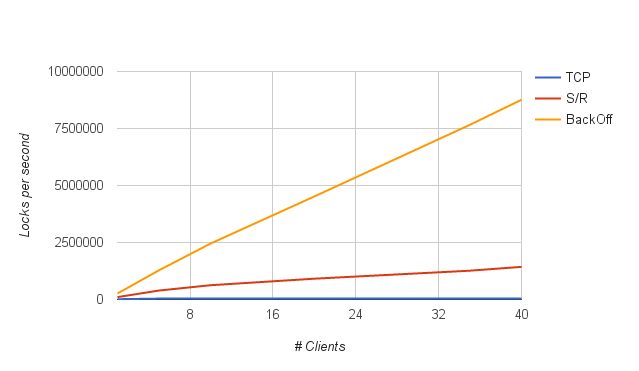}
  \caption{vs. number of client}
\end{subfigure}%
\begin{subfigure}{.5\textwidth}
  \centering
  \includegraphics[width=.8\linewidth]{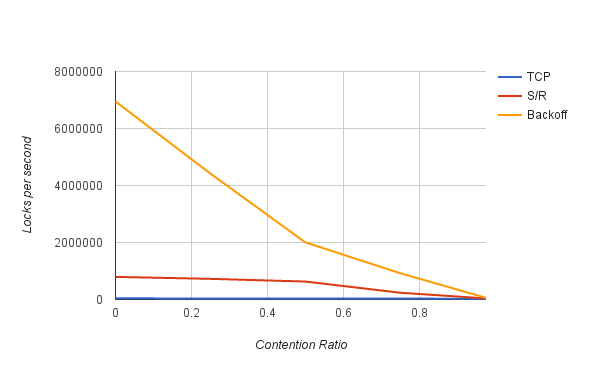}
  \caption{vs. contention rate}
\end{subfigure}
\caption{Throughput comparison results for different designs}
\label{fig:comparison_result}
\end{figure*}

\section{Experiments}
\label{experiments}
We evalute the server centric (TCP, S/R) and the client centric (Backoff) designs for lock operation throughput. The evaluation is done with a small cluster of 7 machines, each equipped with 40-cosre processors.   Figure~\ref{fig:comparison_result} shows the throughput comparison results for server centric and client centric designs. RDMA let clients to bypass server CPU in requesting and receiving locks and the server can handle more lock requests with its limited resources; furthermore, the client centric design outperforms the server centric design, even with RDMA, by a huge margin.


\section{Related Work}
\label{related-works}
Distributed lock management over the traditional network has been the topic of many research works \cite{Aldred:distributed:95, kishida:SSDLM:03}.
\cite{born:analytical:96, knottenbelt:performance:01} presented a detailed analysis of lock management in distributed systems. 
There are a few works which tried to leverage one-sided RDMA verbs to build a DLM \cite{devulapalli:distributed-queue-based-locking:05, narravula:high-performance-distributed-lock-management:07}.
The authors in \cite{devulapalli:distributed-queue-based-locking:05} proposed to use atomic operations to provide.
The basic idea is to have the server only store the tail of the FIFO lists.
Each client keeps track of its immediate subsequent node in the queue.
Once a node is done with a lock, if it has a child, it notifies it by sending an RDMA message to that node.
It is worth noting that their design only supports exclusive locks.
Their techniques were augmented by Narravula et al. \cite{narravula:high-performance-distributed-lock-management:07} to support shared locks. 
Both these papers have the problem of resulting too many connections between clients, and is not scalable with the number of clients.


\section{conclusion}
We set out to remove the bottleneck of the traditional lock manager, which we speculated to be CPU. Using RDMA with the traditional lock manager, we were able to achieve just that; however, re-designing the lock manager and having the clients to partake in lock management can dramatically improve the system throughput. We believe that the proposed design is still not the right way to leverage RDMA for lock management. In the future, we would like to explore several design choices.




\bibliographystyle{abbrv}

\bibliography{main}


\end{multicols}

\end{document}

%% file: server_centric.tex
\section{Server Centric Design}
Centralized lock management is one of the simplest design for a lock service that we often encounter in textbooks. This traditional approach can be easily implemented by allowing a single thread to access a shared memory region or a code section, which effectively serialize concurrent requests from multiple clients. In this section, we first describe the architecture and some important implementation details; next we discuss the performance characteristics of the centralized lock manager over a fast Network (i.e., InfiniBand) using TCP/IP sockets and RDMA send/receive verbs.

A centralized lock manager spawns a thread for each incoming client connection; each thread is responsible for the communication between the server and a connecting client. Messaging via TCP/IP sockets incurs context switching overheads and much of CPU cycles are spent on message handling in kernel. In our experiment with 40 clients, each sending 100,000 lock request and release requests, more than 90\% of server CPU cycles were spent on message handling in kernel.

Any requests received by the server are first registered to an array of FIFO queues; each queue holds lock requests for a specific data item. Any access to a queue within the array is protected by a mutex, and clients will block each other only if they are trying to access the same item. When a lock release request is received by a lock holding client, then the corresponding request is popped from the queue after the server acknowledges the release request; the next requests in the queue is granted the lock. Subsequent shared lock requests can be granted together if no client is holding the same lock in the exclusive mode.

TCP/IP blocking read and write are costly, but the burden can be offloaded to network hardware using RDMA send/receive verbs and the saved CPU cycles can be used to process more lock requests; hence, making the lock manager more scalable.

In Figure~\ref{fig:server_centric_result}(a), the throughput of the traditional design can be improved using RDMA send/receive verbs ($S/R$) and scales with the increasing number of clients. On the other hand, the throughput of the system using TCP/IP socket (TCP) tampers off quickly as the server's CPU is overloaded. With 100 lock items being shared among multiple clients, the server with RDMA tops around at 1.4M locks per second, whereas the one with TCP/IP yields only about 40880 locks per second. Processing a message takes about $10^4$ CPU cycles and assuming forty 3GHz cores, a loose upper bound for the lock throughput is roughly 30M locks per second.

Figure~\ref{fig:server_centric_result}(b) shows how the throughput changes as we vary the contention rate: $CR=1-N_{item}/N_{client}$, where $N_{item}$ is the number of items and $N_{client}$ is the number of clients. Assuming a uniform access pattern for each client, the more clients there are relative to the number of items indicates the higher contention among the clients. Here, the number of clients is fixed at 40.